\documentclass[aps,prd,twocolumn,groupedaddress]{revtex4-1}
\usepackage{graphicx}
\usepackage{subfig}
\usepackage{afterpage}
\usepackage{hyperref}
\usepackage[table]{xcolor}  

\begin{document}
\

\title{Matrix Elements, Parton Showers and Jet Merging: \\ Jet Substructure and New Physics at the LHC}

\author{Andrew Altheimer}
\author{Gustaaf Brooijmans}
\affiliation{Physics Department, Columbia University.}

\date{\today}

\begin{abstract}
Events containing hadronically decaying heavy particles with large momentum, leading to so-called merged jets, are expected to play a significant role in both searches for new physics and measurements of Standard Model processes at the CERN Large Hadron Collider.  In this article, a comparative study of the modeling of such merged jet topologies by different Monte-Carlo event generators is presented.  The observed differences emphasize the need to refine such modeling based on the observation of Standard Model processes prior to claims of discovery.
\end{abstract}


\maketitle

\section{background}
The increased energy and luminosity of modern hadron colliders, first at Run II of the Fermilab Tevatron Collider and now at the CERN Large Hadron Collider (LHC), requires increasingly sophisticated Monte Carlo (MC) generators to model both background and signal processes in order to extract new physics results from the data.  Indeed, one major challenge facing these analyses is that of large final state jet multiplicities, with events containing 6 or more high transverse momentum ($p_T$) jets expected to be common at the LHC.  Additionally, jet structure is likely to play an increasingly important role in physics analyses of hadronic decays of boosted massive particles\cite{Butterworth:2002tt, Djouadi:2007eg, Fitzpatrick:2007qr,Agashe:2006hk, Lillie:2007yh, Brooijmans:2010tn}. The latter aspect in particular probes a difficult region of phase space, testing the ability of MC generators to accurately simulate jet production.

Monte Carlo event generation is commonly performed in two main stages.  A ``matrix element step" evaluates a hard process with a fixed number of incoming and outgoing particles (e.g.~$2 \to 2$), while additional jets may be produced during the parton shower phase of the generator.  While it would be infeasible to evaluate the entire event using the hard process matrix element methods, it is generally difficult to accurately generate multiple additional hard jets in the parton shower phase.  

In order to improve the modeling of events with multiple hard jets, a number of next generation MC generators have been developed to generate hard processes with an increasing number of outgoing particles (up to $2 \to 9$) during the matrix element step.  However, the same jet may now be produced in either the matrix element or parton shower, creating significant ambiguity in the division of phase space for jet production and the potential for ``double counting."  Thus, a proper jet matching algorithm which can dictate which jets should be produced in the hard process and which in the parton shower is crucial to ensure proper phase space coverage, a task which is made difficult by the intrinsic difficulty in linking hard process quarks or gluons to final-state jets.


This article compares the performance of several MC generators.  {\sc Pythia}\cite{Sjostrand:2007gs, Sjostrand:2003wg} and {\sc Herwig++}\cite{Bahr:2008pv} are traditional MC generators which generate only $2\rightarrow2$ hard processes and require any additional jet production to come from the parton shower.  {\sc Alpgen}\cite{ Mangano:2002ea}, a $2 \to n$ ($n \le 9$) generator implements MLM matching, which allows the event evolution to proceed without restriction but afterwards vetoes events whose hard jets do not match the parton-level quarks and gluons produced in the hard process.  CKKW\cite{Catani:2001cc} matching, which suppresses the production of soft jets in the hard process and hard jets during the parton shower phase according to the $k_\perp$ scale of each individual branch splitting as the event is being generated, is implemented in {\sc Sherpa}\cite{Gleisberg:2004hm}, also a $2 \to n$ ($n \le 9$) generator.  

Generators such as {\sc Alpgen} and {\sc Sherpa} offer additional advantages when considering the decays of heavy particles.  While $2 \to 2$ generators can generate heavy particles in the matrix element, heavy particle decays are usually handled at a later phase in the event evolution, before the parton shower.  However, such a factorization is only approximate, and modeling may be improved by incorporating heavy particle decays into the hard process,  as is possible in {\sc Alpgen} and {\sc Sherpa}.  Decay products from heavy particles are not subject to double counting and are generally excluded from the jet matching procedure.  

Hard jets produced in the parton shower or from radiation from the hard process, rather than from the decay of a heavy particle, are hereafter referred to as QCD jets.  In either {\sc Alpgen} or {\sc Sherpa}, the user must specify a maximum number of such QCD jets which may be produced in the hard process.  For events at this upper limit, additional jet production in the parton shower is permitted provided that it is softer than the softest jet from the hard process (according to an algorithm-dependent metric.)  Thus, both of these approaches are designed to produce full phase space coverage without double counting.  

Previous studies of these and other generators have compared the jet $p_T$ and rapidity distributions in $pp$ and $p\overline{p}$ collisions \cite{Alwall:2007fs} as well as $k_\perp$ distance \cite{Lavesson:2007uu} in $e^{+}e^{-}\to jets$ events.  This study  utilizes $pp\to t\overline{t}+jets$ events to examine the structure created through the decay of boosted heavy particles, while high multiplicity $pp\to jets$ (QCD) events provide a more general source of jet mergers.  Both data sets are sensitive to characteristics of the individual matrix element (ME) generators and the parton showers, while the second set in particular may probe differences between CKKW and MLM jet matching.

\section{Jet Merging and Matching Algorithms}

In events with a large jet multiplicity, it will sometimes occur that a pair of `low mass jets' has little separation.  This may result either from the decay kinematics of a boosted heavy particle or from a large multiplicity of uncorrelated jets.  In such cases, jet reconstruction may merge this pair of `low mass jets' into a single jet which acquires a large mass.  This implicitly leads to the appearance of a peak or shoulder in the jet mass distribution whose properties depend on the jet algorithm and its parameters, characteristics of the parton shower, and the kinematics of any heavy particle decays.  Thus, the jet mass distribution provides an experimentally accessible, sensitive probe of the effects of jet merging and the performance of MC generation tools for modeling jet structure.  

In MLM matching, particle-level jets reconstructed with a simple cone algorithm are matched to hard quarks or gluons from the matrix element phase.  In the case that two hard QCD jets are merged by this algorithm, a mismatch between the observed and expected jets will trigger a veto unless exactly one of the jets originates from the ME hard process and the other the parton shower.   On the other hand, hard jets radiated during the parton shower may escape the veto provided they are sufficiently close to a hard process QCD jet so as to become merged (according to the internal cone algorithm.)  However, mapping of hard quarks or gluons to particle-level jets has limited accuracy due to QCD confinement and higher order effects, potentially resulting in improper event vetoes.  

Conversely, in CKKW two jets from the hard process may be allowed to merge without causing a veto.  However, every jet radiated during parton showering must be soft in relation to its point of origin even if it merges with another jet.  Thus, merged hard jets may only originate from the hard process.

In either case, merged jets may be considered to exist along a boundary in phase space between an N and N+1 jet hard process.   Therefore, jet merging occurs in a region of phase space which is particularly sensitive to the phase space coverage produced by jet matching, and the size and position of the shoulder in the jet mass distribution is an excellent probe of the differences between jet matching algorithms (and reality.)

\begin{figure}[!]
\includegraphics[width=\linewidth]{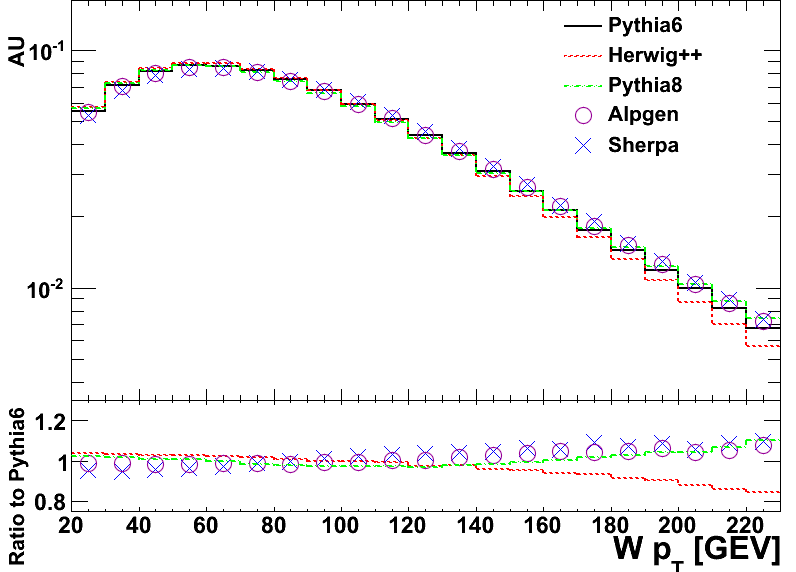}
\caption{\label{fig:Wpt} $W$ boson $p_T$ distribution in $t\overline{t}$ events for several generators.}
\vspace{-10pt}
\end{figure}

\begin{figure*}[!]
\subfloat[D0 Run II Cone, R=1.0] 		{ \includegraphics[width=.33\linewidth]{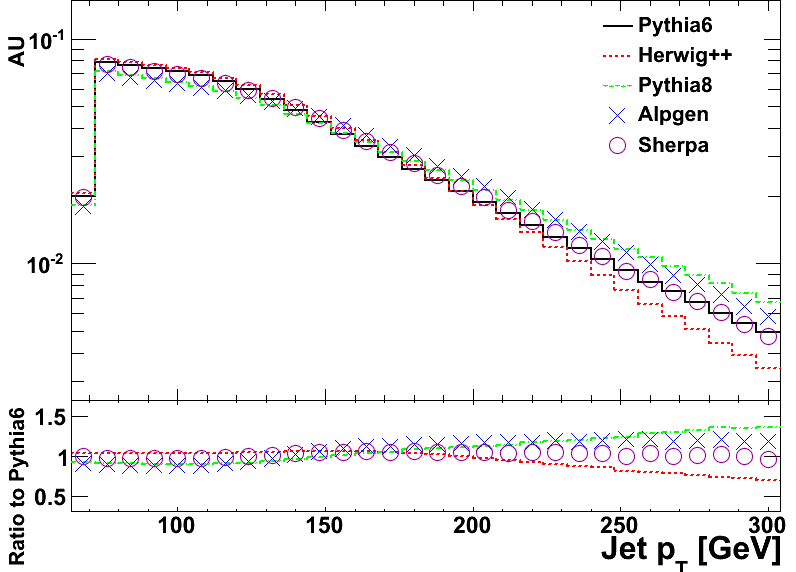}}
\subfloat[Anti-$k_\perp$ R=0.8]			{ \includegraphics[width=.33\linewidth]{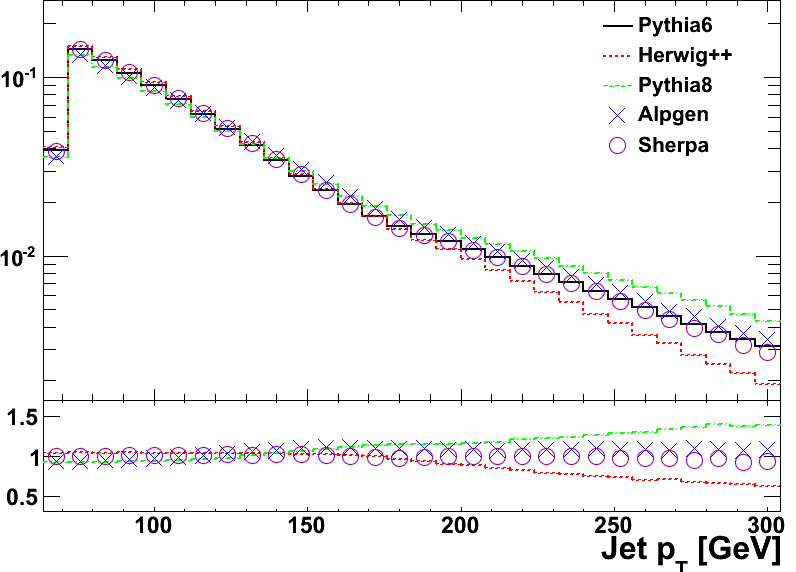}}
\subfloat[Cambridge/Aachen, R=1.0] 	{ \includegraphics[width=.33\linewidth]{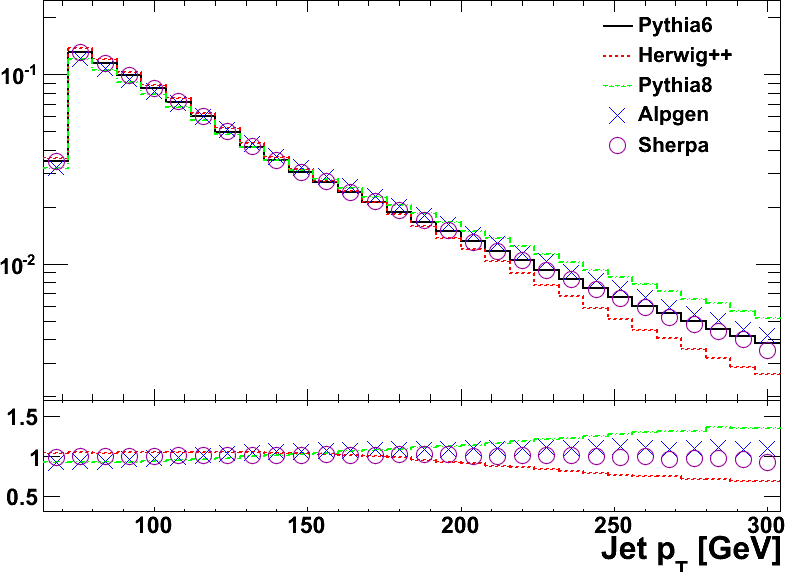}}
\caption{\label{fig:AllPt}$p_T$ distribution of all jets with $|y|<4.5,\  p_T>70$~GeV for several jet clustering algorithms.  Left to right:  D0 Run II Cone R=1.0, anti-$k_\perp$ R=0.8, Cambridge/Aachen R=1.0}
\vspace{-10pt}
\end{figure*}

\begin{figure*}[!]
\subfloat[D0 Run II Cone, R=1.0]		{\includegraphics[width=.33\linewidth]{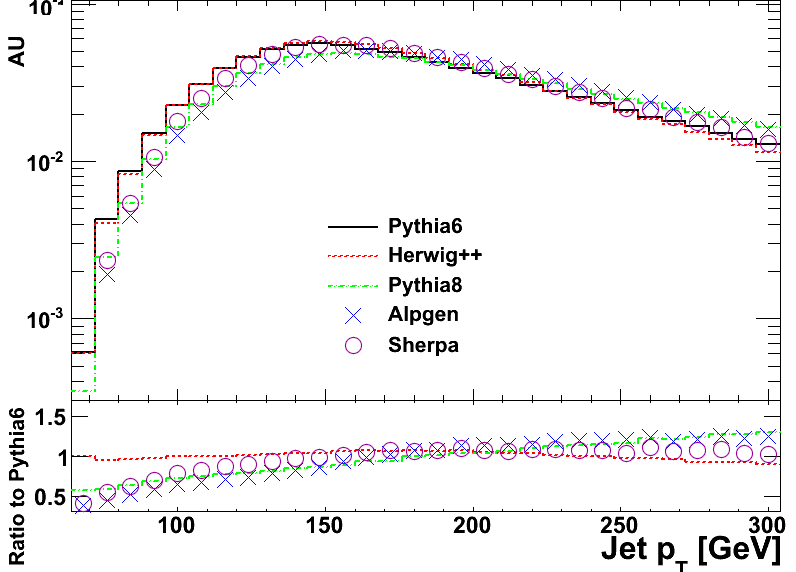}}
\subfloat[Anti-$k_\perp$, R=0.8] 			{\includegraphics[width=.33\linewidth]{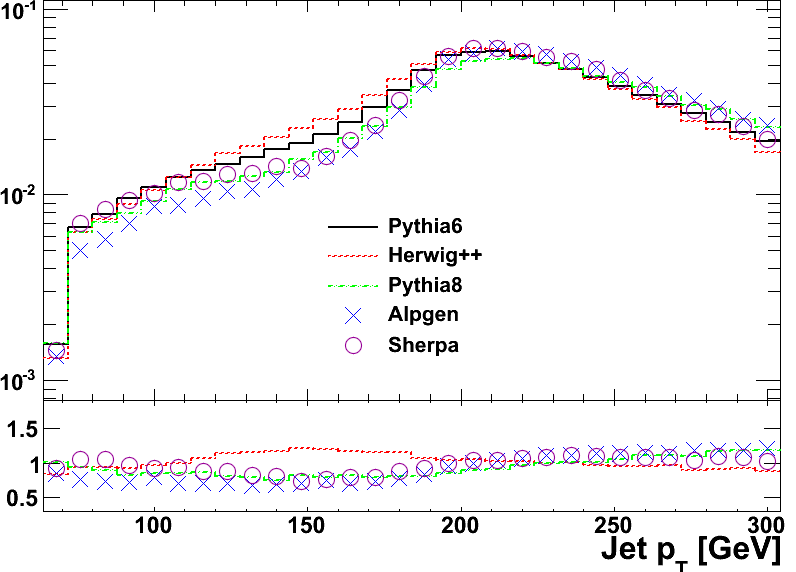}}
\subfloat[Cambridge/Aachen R=1.0]	{\includegraphics[width=.33\linewidth]{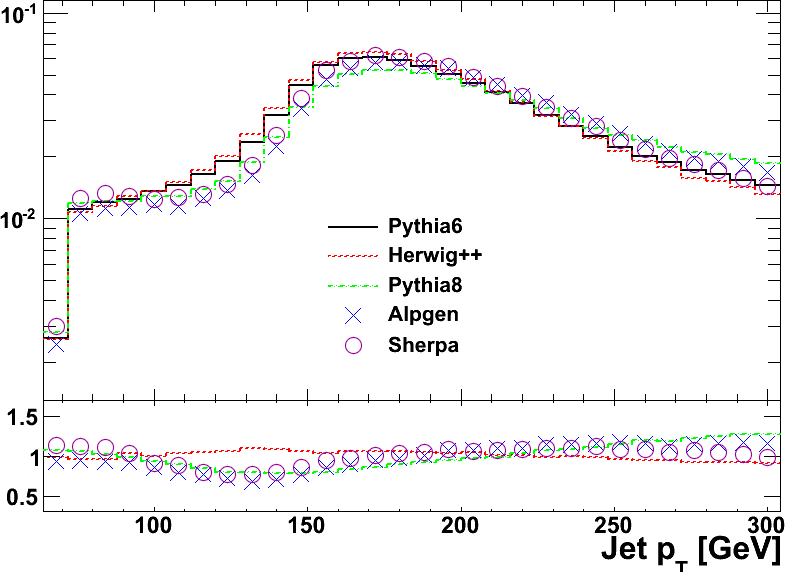}}
\caption{\label{fig:WjetPt}$p_T$ distribution of merged $W$-jets with $|y|<4.5,\ p_T>70$~GeV for several jet algorithms. Left to right:  D0 Run II Cone R=1.0, anti-$k_\perp$ R=0.8, Cambridge/Aachen R=1.0}
\vspace{-10pt}
\end{figure*}

\section{$t\overline{t}$ Event Generation and Jet Reconstruction}
	
	For these studies $pp\to t\overline{t}+jets$ events were generated with several generators at a 14 TeV center-of-mass energy and with the requirement that each top quark decay hadronically into 3 jets via a $W$ boson.  Jet multiplicities of 7-8 were thus achieved, providing a jetty environment in which mergers become likely.  No generator-level cuts were applied and multiple interactions were turned off in every generator.

	The following generators were used:  
\begin{itemize}
\item{{\sc Pythia} 6.325, using the CTEQ5L parton distribution functions (PDFs.)}
\item{{\sc Pythia} 8.150, using the CTEQ5L PDFs.}
\item{{\sc Herwig++} 2.4.2, using the 2008 MRST leading order PDFs.}
\item{{\sc Alpgen} 2.13 for ME generation using the CTEQ5L PDFs and {\sc Pythia} 6.325 for parton showering and hadronization.  Only \mbox{$t\overline{t}$ + 1} (exclusive) or 2 (inclusive) light parton processes were considered, as the production cross section for $t\overline{t}$ + 0 light parton events is vanishingly small.  MLM matching was applied to ensure proper phase space coverage, and the two samples were manually merged according to the cross-section estimates provided by {\sc Alpgen}. By default, {\sc Alpgen} generates only on-shell top quarks and $W$ bosons.}

\item{{\sc Sherpa} 1.2.2 using the COMIX ME generator, which is recommended for large particle multiplicities, and the CTEQ6L PDFs.  The CKKW merging scale was set to 30~GeV.  $t\overline{t}$ + 0, 1, and 2 light parton hard processes were considered and combined internally by the {\sc Sherpa} generator.  Parton showering and hadronization were also performed by {\sc Sherpa}.  Top quarks and $W$ bosons were forced to be on shell.}  

\end{itemize}


	Jets were reconstructed at particle-level (i.e.~after parton shower and hadronization) using the  D0 Run II Cone\cite{Blazey:2000qt}, anti-$k_\perp$\cite{Cacciari:2008gp}, and Cambridge/Aachen\cite{Dokshitzer:1997in} jet algorithms with multiple radii.  This was performed using {\sc Spartyjet} 3.4.1\cite{Ellis:2007ib} and {\sc Fastjet} 2.4.1\cite{Cacciari:2005hq}.  Additionally, parton-level information about the hard process $W$ bosons and their immediate decay products were extracted from the event history.  


\section{$t\overline{t}$ Analysis}

 Figure~\ref{fig:Wpt} shows the particle-level $W$ boson $p_T$ distributions for each generator, with generally good agreement but some divergence at higher $p_T$.  However, Fig.~\ref{fig:AllPt} shows differences in the jet $p_T$ spectrum for all jets with $|y|<4.5,\ p_T>70$~GeV.  


\subsection{Merged $W$-Jets}

\begin{figure*}[!]
\subfloat[D0 Run II Cone, R=1.0]	{\includegraphics[width=.5\linewidth]{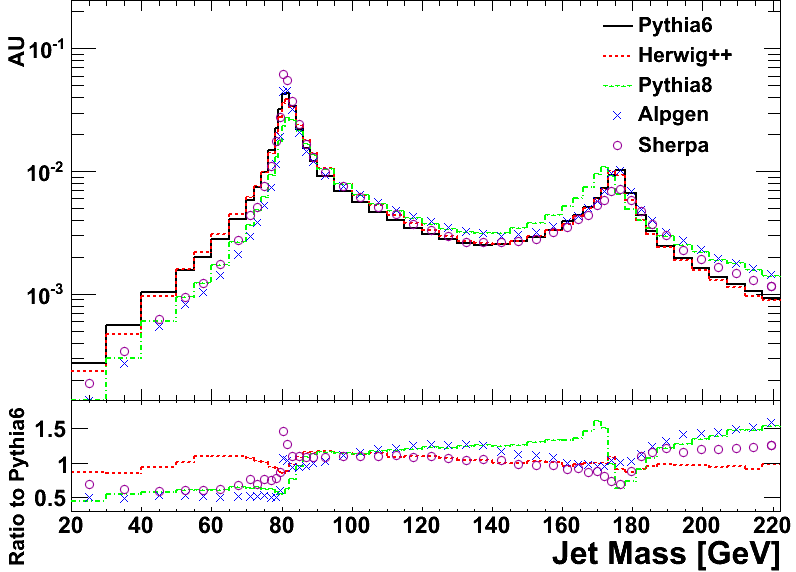}}
\subfloat[Anti-$k_\perp$, R=0.8] 	{\includegraphics[width=.5\linewidth]{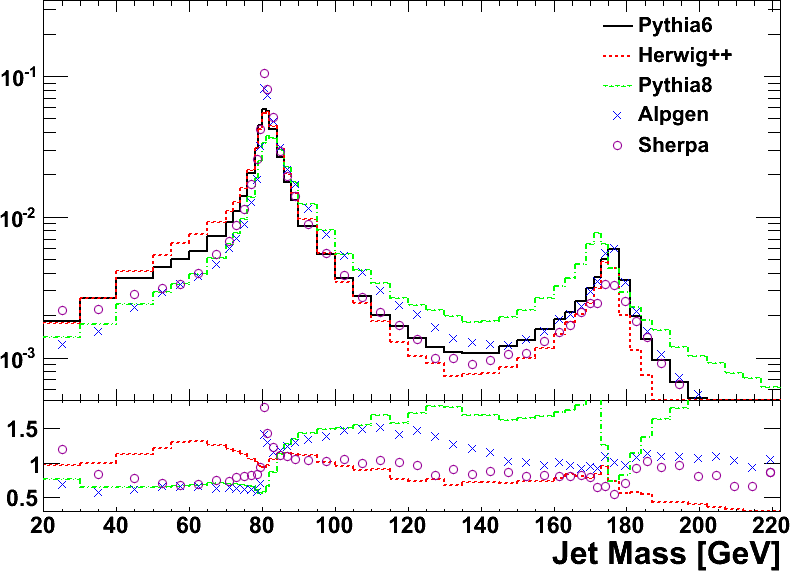}}
\\
\subfloat[Cambridge/Aachen R=1.0]	{\includegraphics[width=.5\linewidth]{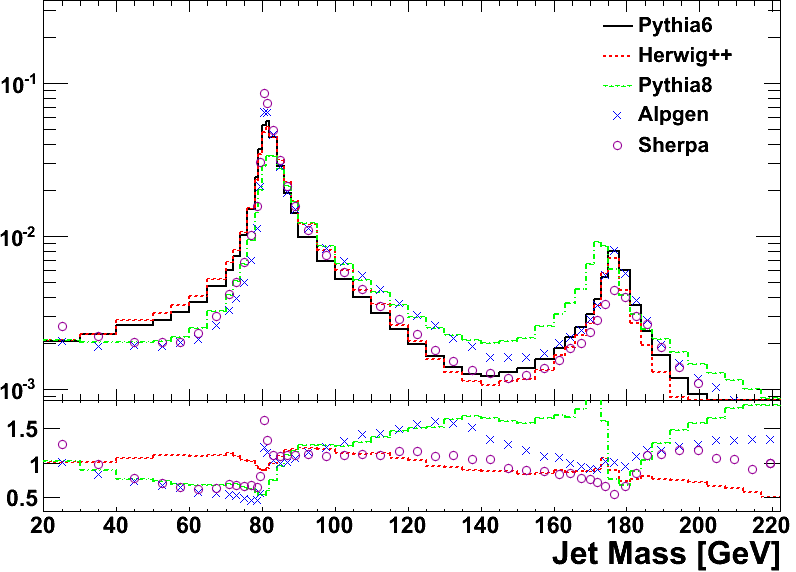}}
\caption{\label{fig:WjetM} Jet mass distribution of merged $W$-jets with $|y|<4.5,\ p_T>70$~GeV for several jet algorithms. Left to right:  D0 Run II Cone R=1.0, anti-$k_\perp$ R=0.8, Cambridge/Aachen R=1.0}
\vspace{-10pt}
\end{figure*}

\begin{figure*}[]
\subfloat[Jet $p_T$] {\includegraphics[width=.50\linewidth]{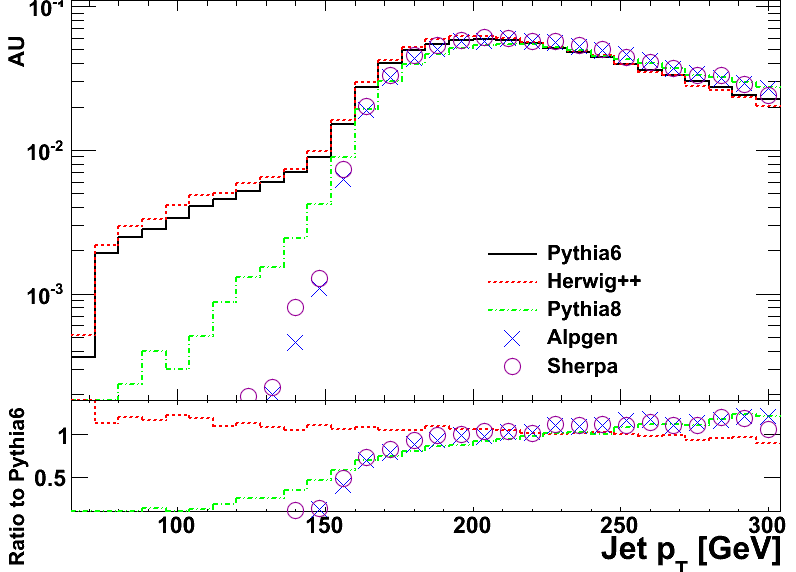}}
\subfloat[Jet mass] {\includegraphics[width=.50\linewidth]{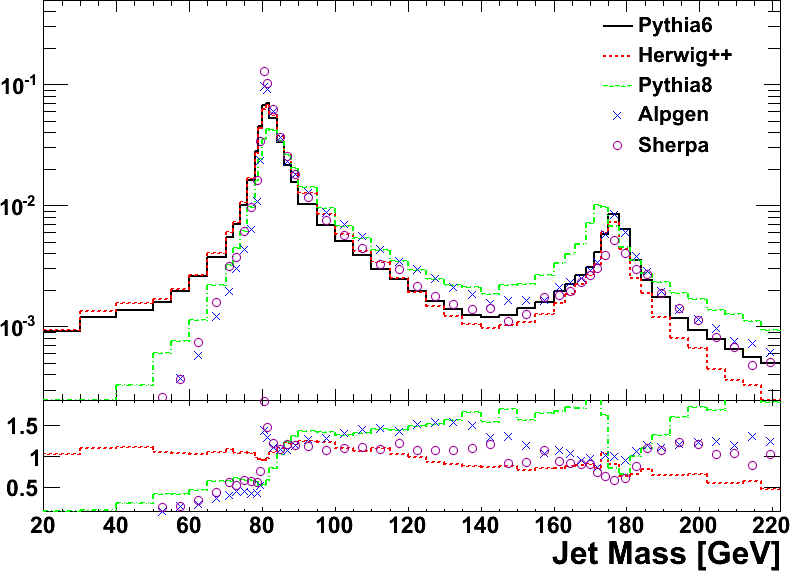}}
\caption{\label{fig:sWjetM} Jet $p_T$ and mass distribution of jets tightly pointed at by a $W$ boson with $y<4.5,\ p_T>70$~GeV.  Both $W$ boson daughter quarks are required to be within R/2 from jet axis.  Jets are reconstructed with Cambridge/Aachen R=1.0}
\vspace{-10pt}
\end{figure*}

\begin{figure*}[]
\subfloat[{\sc Pythia8} vs. {\sc Pythia6}]		{\includegraphics[width=.50\linewidth]{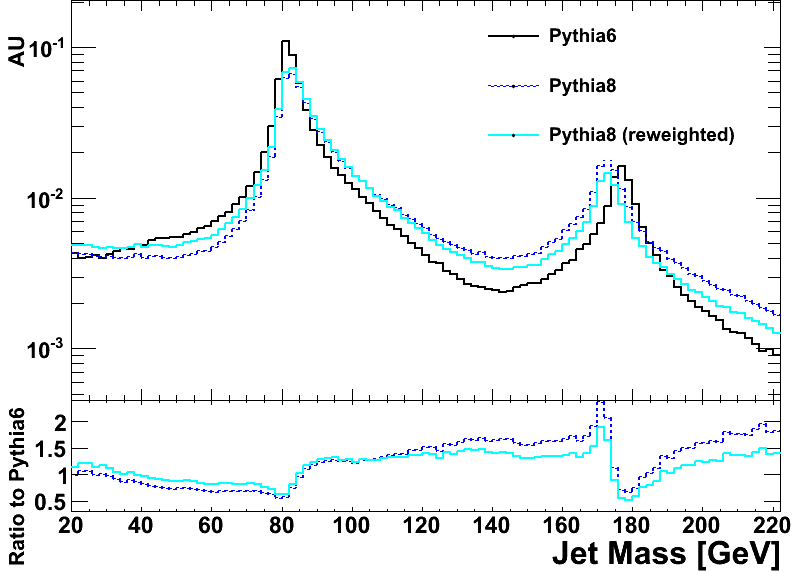}}
\subfloat[{\sc Alpgen} vs. {\sc Pythia6}] 	{\includegraphics[width=.50\linewidth]{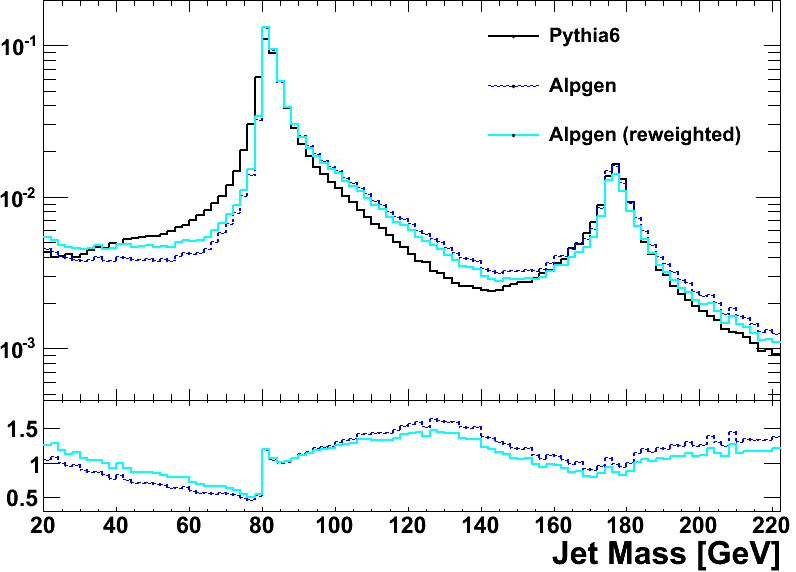}}
\caption{\label{fig:WjetM_r} Jet mass distribution of merged $W$-jets with $|y|<4.5,\ p_T>70$~GeV in {\sc Pythia6}, {\sc Pythia8} (a) and {\sc Alpgen} (b) before and after reweighting jets to match {\sc Pythia6}'s $p_T$ distribution, using Cambridge/Aachen R=1.0 jets.}
\vspace{-10pt}
\end{figure*}

Jets containing both products from a boosted hadronic $W$ boson decay (hereafter referred to as `merged $W$-jets') are considered separately in this study due to their value as an experimental calibration signal for future studies on jet structure.  Additionally, the strong kinematic constraints on such jets generate distinctive features in their mass distribution which must be considered separately from other jet mergers.  Such jets are tagged via the requirement that each of the parton-level decay products from the $W$ boson be sufficiently close to the jet axis ($\sqrt{\Delta y^2+\Delta \phi^2}<R$.)  The remaining bottom quark from the top quark decay is allowed but not required to be contained in the jet as well. 

Figure~\ref{fig:WjetPt} shows a non-trivial difference in the the $p_T$ distribution for merged $W$-jets.  The jet mass distributions shown in Fig.~\ref{fig:WjetM}~display a pair of peaks at approximately 80 and 170~GeV, corresponding to the masses of the $W$ boson and top quark respectively.  

There are several differences between the MC generators which may be seen in these distributions, especially in the treatment of the $W$ boson mass spectra.  {\sc Sherpa} and {\sc Alpgen} force heavy particle decays (both $W$ bosons and top quarks) to be on the mass shell (necessary to keep the computational cost manageable given the increased responsibilities of the hard process.)  This results in a sharper $W$ boson mass peak in {\sc Sherpa} and {\sc Alpgen} than is evident in the $2 \to 2$ generators.  

On the other hand, both {\sc Herwig++} and {\sc Pythia6} generate the $W$ boson according to a relativistic Breit-Wigner mass distribution with a fixed width, while {\sc Pythia8} implements a running width.  This results in a sharp suppression of the low mass tail in {\sc Pythia8} compared to the other $2 \to 2$ generators, and a larger tail in the high mass region.  Although only a fraction of the total, $W$ bosons at the low end of the mass spectrum are more likely to produce merged jets than $W$ bosons which are near or above the mass shell, especially at moderate $p_T$.  This results in a significant excess of `merged $W$ jets' in the low mass low $p_T$ turn on regions in {\sc Herwig++} and {\sc Pythia6} when compared to the other generators.  It is not fully understood whether it is better to use a fixed or running mass width in these decays, and these differences should be viewed as a systematic uncertainty in the modeling of top quark decays and other heavy particles\cite{pythiaComm}.  

However, differences in particle width modeling do not appear to have as strong of an effect on the shape of the second mass peak, as {\sc Sherpa} and {\sc Alpgen} do not show narrower distributions than the other generators despite generating top quarks on the mass shell.  Rather, in this case {\sc Pythia6}, {\sc Herwig++} and {\sc Alpgen}, all of which use virtuality ordered parton showers, show close agreement in the shape of this peak when compared to {\sc Pythia8} and {\sc Sherpa} which shower in $p_T$ and $k_\perp$ respectively.  

It is important to note that the generators may show differences in the documentation of intermediary particles in the event history which may influence the tagging of merged $W$-jets.  Since such intermediary particles are inherently unphysical, these differences should not be reflected in the physics predictions produced by the generators.  However, it is possible that they would influence the tagging criteria in this analysis.  Figure \ref{fig:sWjetM} shows the effect of tightening the selection criteria on merged $W$-jets by requiring both quarks to be within \emph{half} the jet radius.   The relative insensitivity of these distributions to the tagging criteria suggests that the observed disagreements reflect differences in the final states produced by the generators, rather than internal differences in the event record.  This tightening does however suppress the low mass, low $p_T$ end of the distributions, which in turn makes the influence of the $W$ boson width modeling more obvious.  

Differences in the jet $p_T$ spectrum between generators account for only a small part of the differences in the mass spectrum.  Reweighting the merged $W$-jets such that the jet $p_T$ distributions match does lead to some improve the agreement between generators but does not fundamentally alter the behavior, as shown in Fig.~\ref{fig:WjetM_r}.  

\begin{figure*}[]
\subfloat[{\sc Pythia8}]		{\includegraphics[width=.33\linewidth]{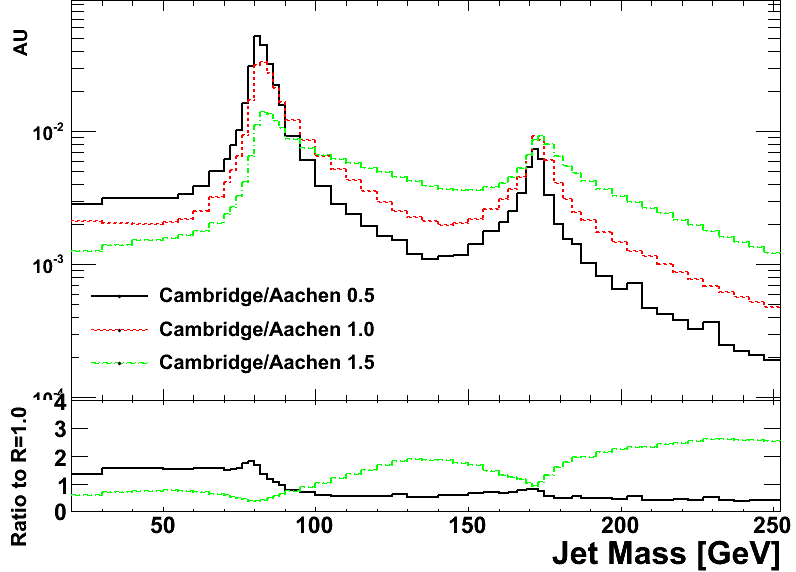}}
\subfloat[{\sc Alpgen}] 		{\includegraphics[width=.33\linewidth]{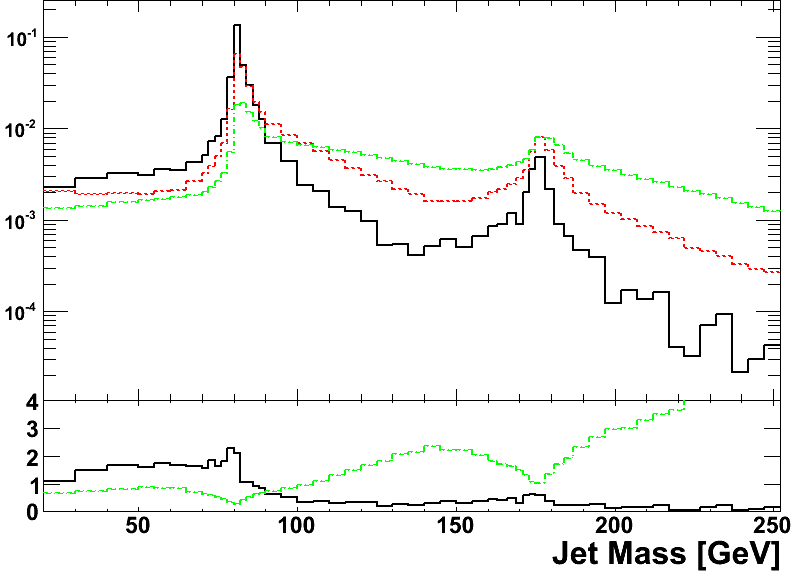}}
\subfloat[{\sc Sherpa}] 		{\includegraphics[width=.33\linewidth]{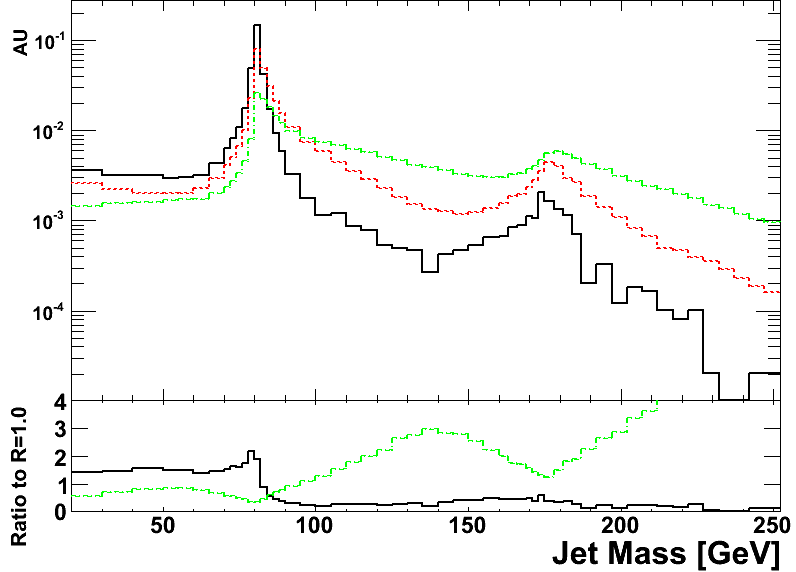}}
\caption{\label{fig:Wradii_Mass} Jet mass distribution of merged $W$-jets with $|y|<4.5,\ p_T>70$~GeV in {\sc Pythia8}, {\sc Alpgen} and {\sc Sherpa} for several jet radii.}
\vspace{-5pt}
\end{figure*}

Increasing the jet radius suppresses the first peak in the mass spectrum and fills in the gap between the two peaks, suggesting that it becomes more difficult to create a `merged $W$-jet' which does not include at least some additional radiation, perhaps from the third jet. This is shown in Fig.~\ref{fig:Wradii_Mass} for {\sc Pythia 8}, {\sc Alpgen} and {\sc Sherpa}.  Differences in behavior between the generators as the jet radius is varied would suggest that different generators may lead to different jet radius optimizations for the tagging of heavy particle decays.

\begin{figure*}[]
\subfloat[D0 Run II Cone, R=1]		{\includegraphics[width=.33\linewidth]{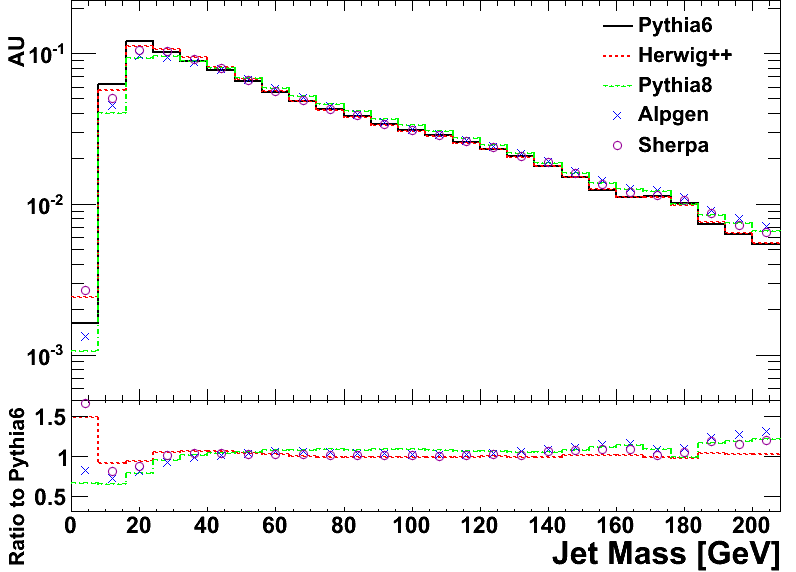}}
\subfloat[Anti-$k_\perp$, R=.8] 			{\includegraphics[width=.33\linewidth]{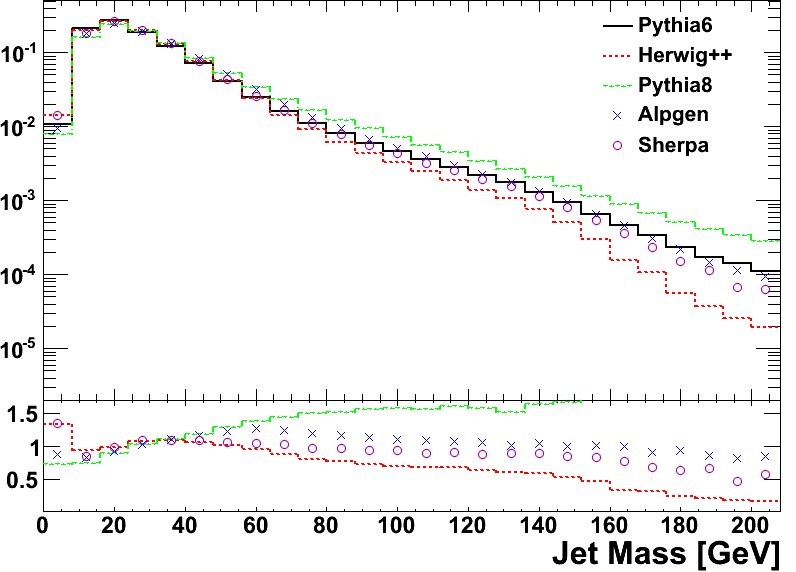} \label{fig:NoWjetM_b}}
\subfloat[Cambridge/Aachen, R=1]	{\includegraphics[width=.33\linewidth]{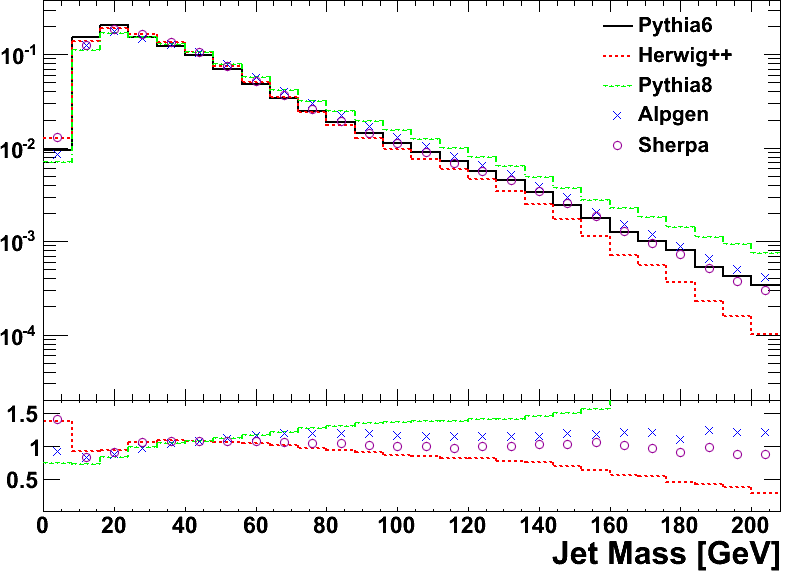}}
\caption{\label{fig:NoWjetM} Mass distribution of jets not pointed at by a $W$ boson with $|y|<4.5,\ p_T>70$~GeV for several jet clustering algorithms.  Left to right:  D0 Run II Cone R=1.0, anti-$k_\perp$ R=1.0, Cambridge/Aachen R=1.0} 
\vspace{-5pt}
\end{figure*} 

\subsection{Bottom Jet Mergers}

\begin{figure}[]
\includegraphics[width=\linewidth]{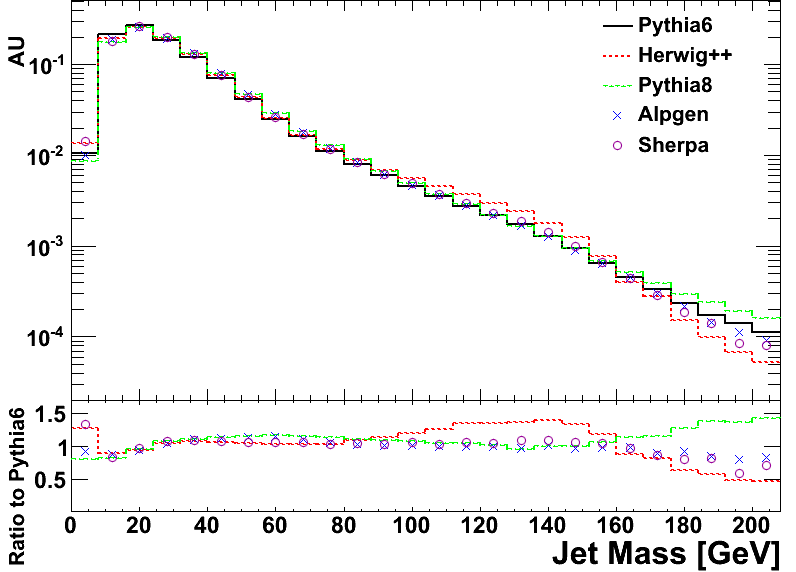}
\caption{\label{fig:NoWjetM_r} Mass distribution of jets not pointed at by a $W$ boson with $|y|<4.5, p_T>70$~GeV for the anti-$k_\perp$ R=0.8 jet algorithm after jet $p_T$ reweighting.}
\vspace{2pt}
\end{figure}

\begin{figure}[]
\includegraphics[width=\linewidth]{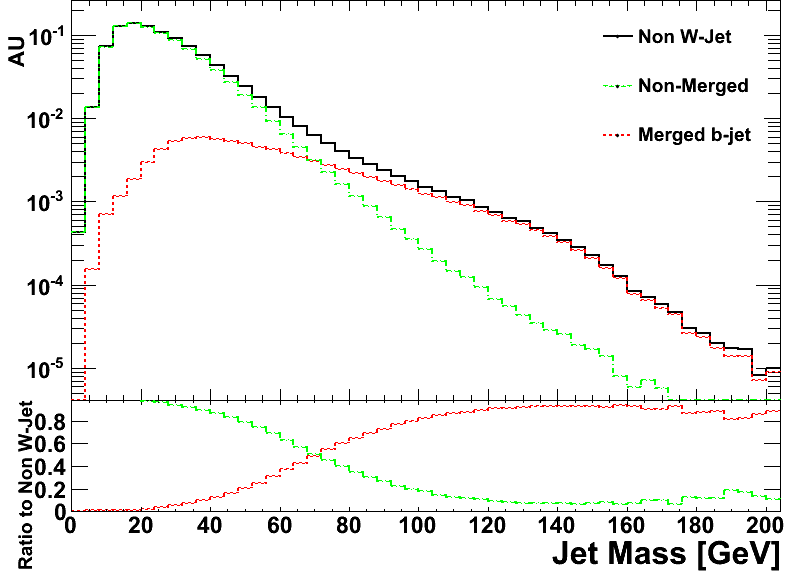}
\caption{\label{fig:bjet} Mass distribution of jets not pointed at by a $W$ boson with $|y|<4.5, p_T>70$~GeV for the anti-$k_\perp$ R=0.8 jet algorithm.  Merged b-jets are those which are pointed at by a bottom quark and exactly one $W$ boson decay product from the same top quark.  Non-merged jets include all jets which are neither merged W-jets nor merged b-jets (but may still contain jet mergers).  Shown for {\sc Herwig++}.}
\vspace{-10pt}
\end{figure} 

\begin{figure*}[]
\subfloat[]  {\includegraphics[width=.50\linewidth]{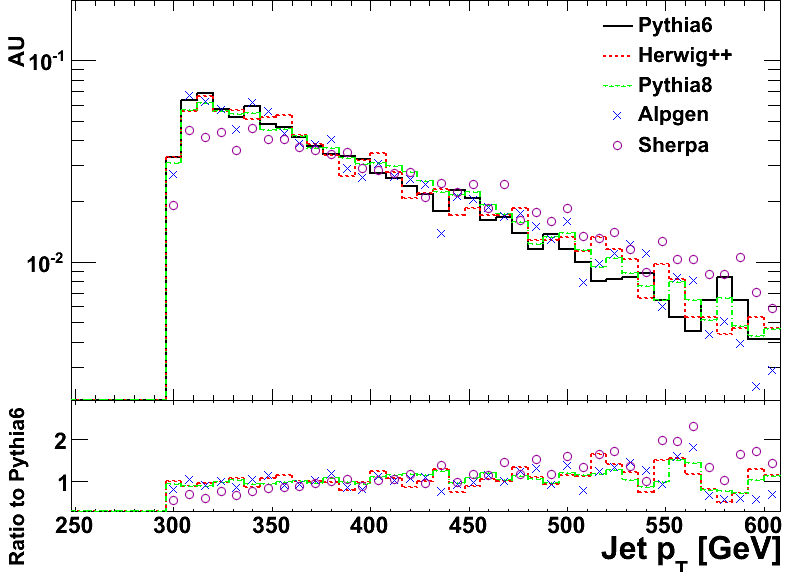}}
\subfloat[]  {\includegraphics[width=.50\linewidth]{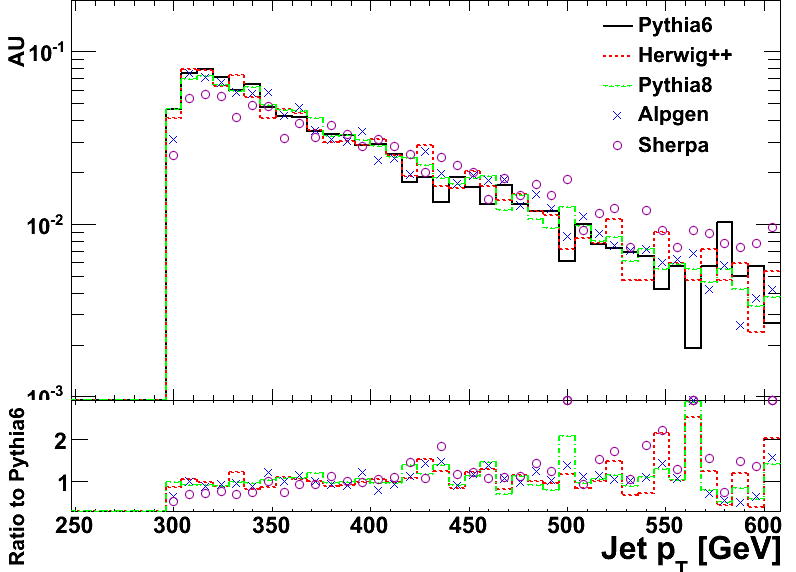}}
\\
\subfloat[]  {\includegraphics[width=.50\linewidth]{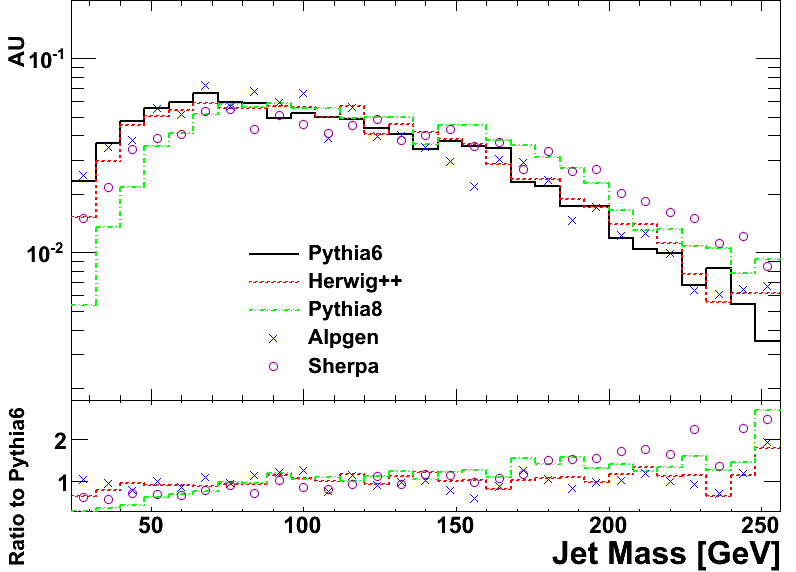}}
\subfloat[]  {\includegraphics[width=.50\linewidth]{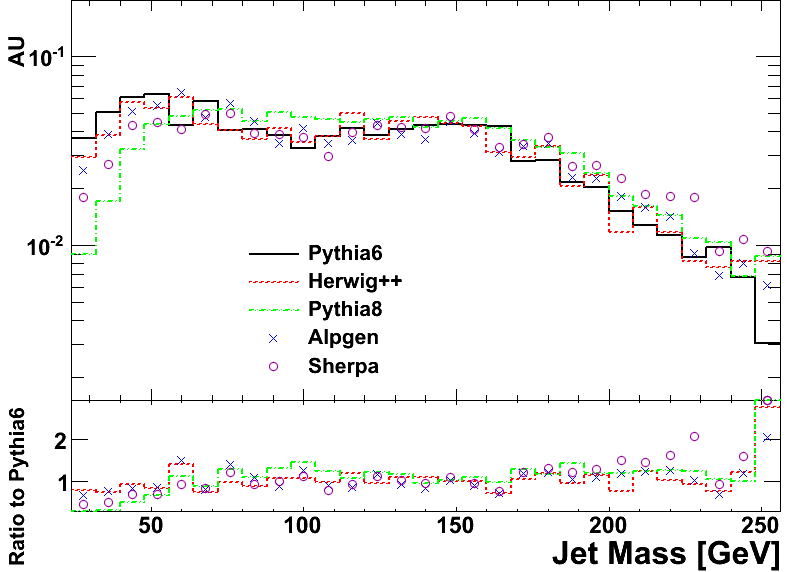}}
\caption{\label{fig:QCDlead} $p_T$ and mass distribution of the leading jet ($|y|<1.5, p_T>300$~GeV) reconstructed with the anti-$k_\perp$, R=1.0 jet algorithm for QCD multijet events with multiple `narrow' jets in a tight range $|y|<1.5$.  Narrow jet preselection is based on either anti-$k_\perp$ (a,c) or D0 Run II Cone (b,d) with R=0.2.}
\vspace{0pt}
\end{figure*} 

\begin{figure}[]
\includegraphics[width=\linewidth]{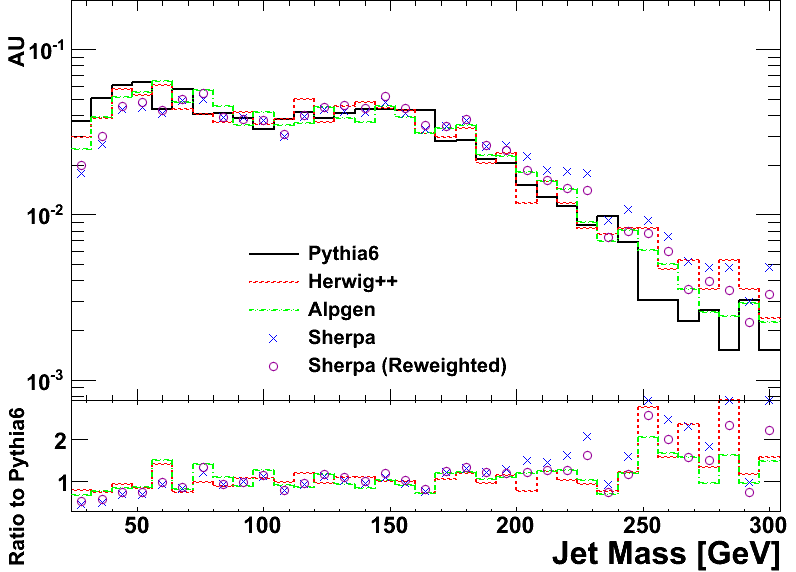}
\caption{\label{fig:QCDlead_r} Mass distribution of the leading jet ($|y|<1.5, p_T>300$~GeV) reconstructed with the anti-$k_\perp$, R=1.0 jet algorithm for QCD events with multiple `narrow' (D0 Run II Cone R=.2) jets.  Shown for the Sherpa MC generator, before and after reweighting by leading jet $p_T$.}
\vspace{0pt}
\end{figure} 

\begin{figure*}[]
\subfloat[]  {\includegraphics[width=.50\linewidth]{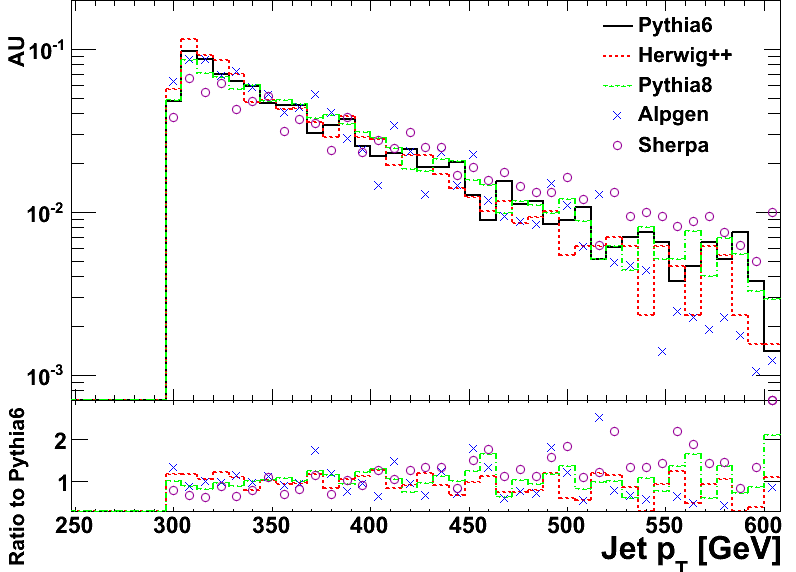}}
\subfloat[]  {\includegraphics[width=.50\linewidth]{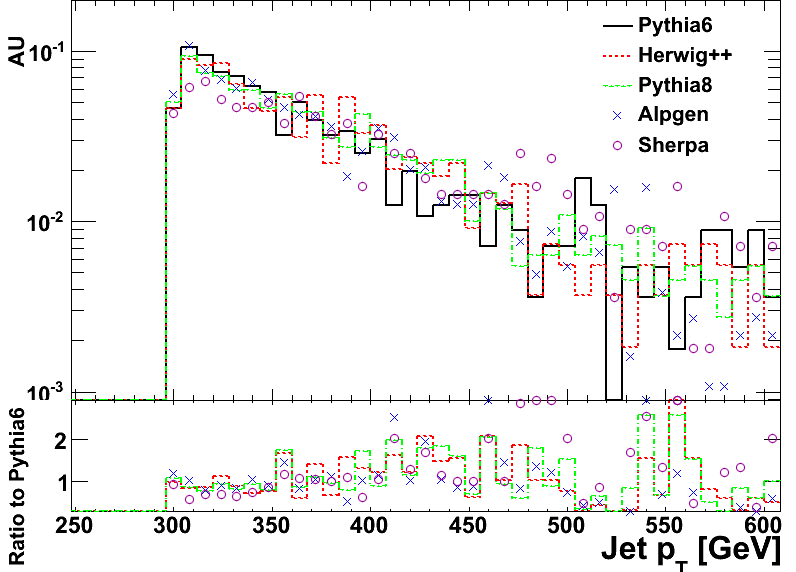}}
\\
\subfloat[]  {\includegraphics[width=.50\linewidth]{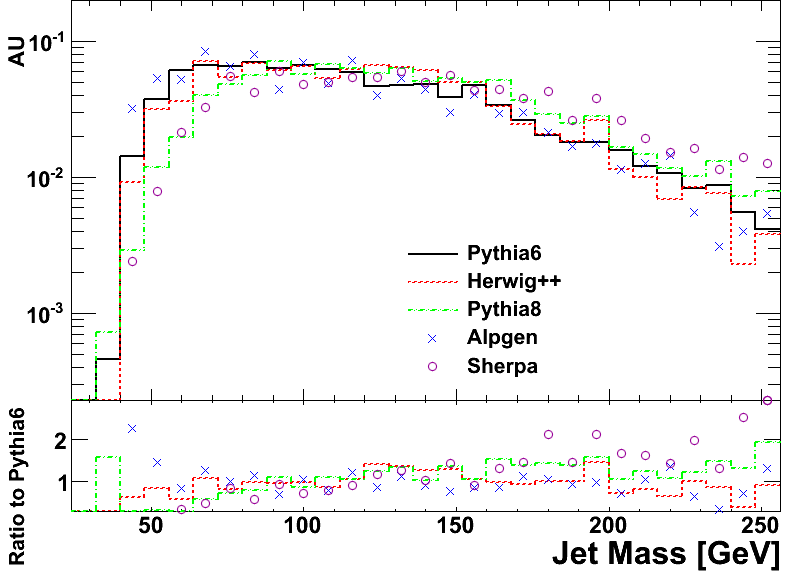}}
\subfloat[]  {\includegraphics[width=.50\linewidth]{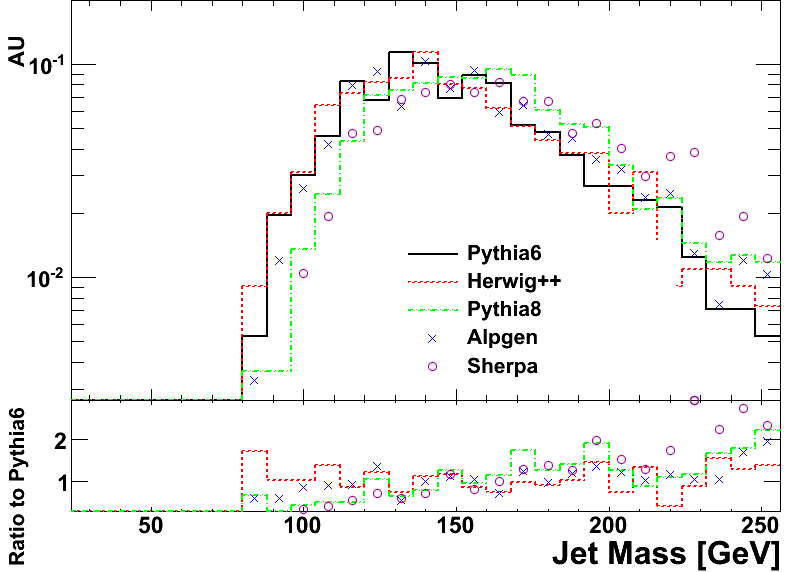}}
\caption{\label{fig:QCD1lead} $p_T$ and mass distribution of the leading jet ($p_T>300$~GeV, $|y|<1.5$) in QCD multijet events, reconstructed with the anti-$k_\perp$, R=1.0 jet algorithm and requiring that multiple 'narrow' jets be contained within the jet radius.  Narrow jet preselection is based on either anti-$k_\perp$ (a,c) or D0 Run II Cone (b,d), R=0.2.}
\vspace{0pt}
\end{figure*}

Merged jets which contain the majority of the radiation produced in the decay of a single heavy particle (such as a top quark or $W$ boson) are subject to tight kinematic constraints.  However, it is also possible that the bottom quark resulting from a top quark decay will merge with exactly one of the $W$ boson decay products from the same top quark (hereafter referred to as `merged $b$-jets.')  Such mergers are less constrained by decay kinematics, limiting their usefulness as a calibration tool.  However, such mergers may prove to be a non-trivial background in other physics analyses.    

Figure \ref{fig:NoWjetM} shows the mass distributions for all jets excluding merged $W$-jets and requiring $p_T>70$~GeV, $|y|<4.5$ for each generator and several jet algorithms.  The small bump in the distribution produced by the D0 Run II Cone algorithm at around 170~GeV suggests some inefficiency in the tagging of such mergers, possibly because this algorithm absorbs nearby radiation easily, producing a larger effective jet area for a given radius.  Additionally, the cone algorithm appears to be less sensitive to differences between the generators in this and several previous distributions.  

There is a small shoulder in these mass distributions appearing between 100 and 120~GeV, especially for the anti-$k_\perp$ algorithm.  This shoulder appears to share a similar position for each generator, and is most apparent in those generators with the softest jet mass spectra.  The disagreement between generators in these distributions is at least partially explained by differences in the $p_T$ spectrum.  This is demonstrated in Fig.~\ref{fig:NoWjetM_r}, which shows the mass distributions for non-`merged $W$ jets' after these jets are reweighted such that the jet $p_T$ distribution matches {\sc Pythia6}.  However, after such a reweighting is applied, the mass shoulder appears to be more significant in those generators with the softest original jet mass spectrum, especially {\sc Herwig++}.

Figure ~\ref{fig:bjet} shows the effect of separating out `merged b-jets' from the entire non-`merged $W$ jet' distribution.  This shows that the shoulder in the mass distribution is primarily produced by mergers between a bottom quark and the $W$ boson decay product from the same top quark.

\section{QCD Event Generation and Reconstruction}

The preceding sections discussed the production of merged jets which are produced in the decay of boosted heavy particles.  The following two sections will explore more general jet mergers which may be produced in any event with a large jet multiplicity.  

For this purpose, QCD multijet events were generated using the same set of generators as described in section III ({\sc Pythia 8.135} was used instead of 8.150.)  Only weak generator-level cuts were applied in order to minimize the introduction of bias, as cuts available to $2\to2$ generators are generally not equivalent to those available for $2\to n$ processes.  In each $2\to2$ generator, $p_T>100$~GeV was required in the hard process.  The following generator settings were used in {\sc Alpgen} and {\sc Sherpa}:

\begin{itemize}
\item{{\sc Alpgen}: 2-5 jet hard processes were generated with the requirement that all partons have $p_T>40$~GeV, $|y|<2.5$.  After parton showering, MLM matching was applied at a scale of 48~GeV and requiring jet $|y|<2.0$. }
\item{{\sc Sherpa}: 2-5 jet hard processes were generated using the COMIX ME generator, with the CKKW merging scale at 30~GeV.  The two primary parton-level jets in the event are required to have $p_T>60$~GeV.  The leading particle-level anti-$k_\perp$ $R=1.3$ jet in the event with $|y|<2.0$ is required to have $p_T>200$~GeV.}
\end{itemize}

As in section III, multiple interactions were turned off in every generator and jets were reconstructed using several jet algorithms.

\section{QCD Analysis}

This analysis of QCD events proceeds by selecting events with a large multiplicity of `narrow' (small radius) jets before applying a `fat' (larger radius) jet algorithm and observing the resulting mass distribution.  The performance of the MC generators is then evaluated by observing the degree of agreement between them.  In this section, two variations of this analyses will be presented.

In the first variation, events are preselected if they contain at least four `narrow' jets (ie.~jets which are reconstructed with either the D0 Run II Cone or anti-$k_\perp$ algorithm with R=.2) with $|y|<1.5$ and $p_T>60$~GeV, and at least one with $p_T>170$~GeV.  Events are then required to have a leading (anti-$k_\perp$ with R=1.0) jet with $p_T>300$~GeV and $|y|<1.5$.  

Figure~\ref{fig:QCDlead} shows the leading jet $p_T$ and mass distributions of events surviving this selection.  The mass distribution produced by the D0 Run II Cone narrow jet preselection shows a clear shoulder produced by the presence of merged narrow jets, while the distribution produced by anti-$k_\perp$ preselection suggests a similar, although less pronounced effect.  

{\sc Pythia8} appears to show some disagreement with the other generators, with a slightly slower turn on under either choice of preselection `narrow' jet definition.  This has the effect of reducing the apparent dip between the two mass peaks when the D0 Run II Cone narrow jet algorithm is used, and creating a more apparent shoulder when the anti-$k_\perp$  narrow jet preselection is used.  {\sc Sherpa} appears to have a harder spectrum than the other generators, and Fig.~\ref{fig:QCDlead_r} shows that reweighting {\sc Sherpa}'s $p_T$ spectrum to match {\sc Herwig++} improves the agreement in the mass distribution.

A second analysis technique leads to a more direct selection of merged jets.  This is achieved by selecting a leading fat (anti-$k_\perp$ with R=1.0) jet with $|y|<1.5$ and $p_T>300$~GeV and requiring at least 2 narrow jets with $p_T>60$~GeV be contained within its jet radius.  The result of this approach can be seen in Fig.~\ref{fig:QCD1lead}, which shows the leading jet $p_T$ and mass distributions for both narrow jet definitions.  

Such `merged QCD' jets form a nontrivial fraction of all leading QCD jets.  Table~\ref{effTable} displays the efficiency for events in each generator which contain at least one fat jet with $p_T>300$~GeV and $|y|<1.5$ to survive these selection cuts.  Furthermore, although the shapes of the distributions shown in Fig.~\ref{fig:QCD1lead} are similar in each generator, the efficiencies vary significantly, highlighting the need to better understand the production of such QCD jets as a potential background in other analyses utilizing jet structure.   

\begin{table*} 
\begin{tabular}{|c|c|c|c|c|}
\hline
			&   \multicolumn{2}{c|} {Selection 1} 			&       \multicolumn{2}{c|}{ Selection 2} 			\\    	\hline
narrow jet 	:	& \ \ \ \ \ \ Anti-$k_\perp$\ \ \ \ \ \  &  D0 Run II Cone		& \ \ \ \ \ \ Anti-$k_\perp$\ \ \ \ \ \ 	&  D0 Run II Cone            	\\	\hline
{\sc Pythia6}     	&	$.137 \pm .002$			&	$.070 \pm .001$		&	$.048 \pm .001$		&	$.019 \pm .001$		\\	
{\sc Herwig++}	&	$.126 \pm .002$			&	$.066 \pm .002$		&	$.051 \pm .001$		&	$.021 \pm .001$		\\	
{\sc Alpgen}      	& 	$.133 \pm .003$			& 	$.071 \pm .001$		&	$.045 \pm .002$		&	$.015 \pm .001$		\\	
{\sc Pythia8}	  	&	$.163 \pm .002$			&	$.128 \pm .002$		&	$.078 \pm .001$		&	$.038 \pm .001$		\\	
{\sc Sherpa}      	&	$.181 \pm .002$			&	$.109 \pm .002$		&	$.057 \pm .001$		&	$.028 \pm .001$		\\	\hline

\end{tabular}{}
\caption{\label{effTable} Efficiency for events with at least one jet with $p_T>300$~GeV and $|y|<1.5$ to survive event preselection criteria for each generator.}

\end{table*}

\section{Conclusions}

This study has examined the modeling of merged jets both from boosted particle decays and from QCD multijet backgrounds.  Such jet mergers provide a window into MC jet modeling and are a sensitive probe of jet matching, parton showering, and other characteristics of MC generators.  While decent overall agreement exists between the generators, significant differences, both explained and unexplained, were observed in key observables such as jet mass, with substantial variations in event rates observed in several regions of phase space.  Such differences may provide a handle towards a better understanding of systematic uncertainties which may affect precision measurements and searches for new physics at the LHC, especially those involving the decay of boosted heavy particles, and may prove to be a useful starting point for future studies to improve the performance of modern MC generators by comparing these predictions to LHC data as it becomes available.

\begin{acknowledgments}

We thank the participants of the Joint Theoretical-Experimental Workshop on Jets and Jet Substructure at the LHC held at the University of Oregon in January, 2011 and supported in part by the DOE under Task TeV of contract DE-FG02-96ER40956 for useful discussions.  We also thank T. Sj\"ostrand and P. Skands for their aid in understanding key differences between {\sc Pythia6} and {\sc Pythia8}.  This work was done thanks to the support of the National Science Foundation.

\end{acknowledgments}

\bibliography{jetMerge}

\end{document}